\documentclass[%
reprint,
amsmath,amssymb,
aps,
prl,
noeprint,
nolongbibliography
]{revtex4-2}

\usepackage{graphicx}
\usepackage{dcolumn}
\usepackage{bm}
\usepackage{xcolor}
\usepackage{amsmath}
\usepackage{lineno}
\usepackage{overpic}
\usepackage{natbib}
\usepackage{hyperref}
\usepackage[utf8]{inputenc}
\hypersetup{colorlinks=true, citecolor=blue, urlcolor=blue, linkcolor=blue}

\newcommand{\Lambdabar}{\bar{\Lambda}}
\newcommand{\mbeq}{\overset{!}{=}}
\usepackage{hyperref}
\begin{document}

\title{\boldmath Unraveling the structure of $\Lambda$ hyperons with polarized $\Lambda\bar{\Lambda}$ pairs}

\author{
M.~Ablikim$^{1}$, M.~N.~Achasov$^{4,c}$, P.~Adlarson$^{76}$, X.~C.~Ai$^{81}$, R.~Aliberti$^{35}$, A.~Amoroso$^{75A,75C}$, Q.~An$^{72,58,a}$, Y.~Bai$^{57}$, O.~Bakina$^{36}$, Y.~Ban$^{46,h}$
% ...
\\
\vspace{0.2cm}
(BESIII Collaboration)\\
\vspace{0.2cm} {\it
$^{1}$ Institute of High Energy Physics, Beijing 100049, People's Republic of China\\
% ...
$^{4}$ Budker Institute of Nuclear Physics SB RAS (BINP), Novosibirsk 630090, Russia\\
% ...
$^{35}$ Johannes Gutenberg University of Mainz, Johann-Joachim-Becher-Weg 45, D-55099 Mainz, Germany\\
$^{36}$ Joint Institute for Nuclear Research, 141980 Dubna, Moscow region, Russia\\
% ...
$^{46}$ Peking University, Beijing 100871, People's Republic of China\\
% ...
$^{57}$ Southeast University, Nanjing 211100, People's Republic of China\\
$^{58}$ State Key Laboratory of Particle Detection and Electronics, Beijing 100049, Hefei 230026, People's Republic of China\\
% ...
$^{72}$ University of Science and Technology of China, Hefei 230026, People's Republic of China\\
% ...
$^{75}$ University of Turin and INFN, (A)University of Turin, I-10125, Turin, Italy; (B)University of Eastern Piedmont, I-15121, Alessandria, Italy; (C)INFN, I-10125, Turin, Italy\\
$^{76}$ Uppsala University, Box 516, SE-75120 Uppsala, Sweden\\
% ...
$^{81}$ Zhengzhou University, Zhengzhou 450001, People's Republic of China\\

\vspace{0.2cm}
$^{a}$ Deceased\\
% ...
$^{c}$ Also at the Novosibirsk State University, Novosibirsk, 630090, Russia\\
% ...
$^{h}$ Also at State Key Laboratory of Nuclear Physics and Technology, Peking University, Beijing 100871, People's Republic of China\\
% ...
}
}

\date{\today}

\begin{abstract}
With data collected in a dedicated energy scan from 2.3864 up to 3.0800 GeV, the BESIII collaboration provides the first complete energy-dependent measurements
of the $\Lambda$ electromagnetic form factors in the time-like
region. By combining double-tag and single-tag events from the $e^+e^-
\to \Lambda \Lambdabar \to p\pi^-\bar{p}\pi^+$ reaction, we achieve a
complete decomposition of the spin structure of the reaction at
five energy points, with high statistical and systematic
precision. Our data reveal that while the modulus of the ratio between
the electric and magnetic form factor, $R(q^2) = |G_E(q^2)/G_M(q^2)|$,
remains fairly constant across the considered energy range, the
relative phase $\Delta \Phi(q^2) = \Phi_E(q^2) - \Phi_M(q^2)$ changes
by more than 90$^{\mathrm{o}}$ between 2.396 and 2.6544 GeV. Using a
fit to our data based on dispersion relations, the complex form factor
ratio is determined as a function of $q^2$ and the preferred solution has multiple zero-crossings in the complex plane. From the derivative of the ratio at $q^2 = 0$, the
root-mean-squared charge radius of the $\Lambda$ is obtained. The two
most probable solutions yield a negative root-mean-squared charge
radius, indicating an asymmetric charge distribution where the $ds$
quark pair lies close to the center of the $\Lambda$ hyperon.

\end{abstract}

\maketitle 
To understand how almost massless quarks form massive
hadrons is a challenging endeavor, involving the full complexity of
the strong interaction and its emergent phenomena. Electromagnetic
form factors (EMFFs) provide a powerful tool to study the strong
interaction. Probed in virtual photon-hadron processes, EMFFs quantify
hadron structure in terms of analytic functions of the squared
four-momentum transfer $q^2$ carried by the photon. More than 70 years
of detailed investigations employing elastic electron-proton
scattering have unraveled the spatial distribution and motion of
quarks and gluons inside the proton \cite{Punjabi:2015bba,
  Pacetti:2014jai}. In particular, the electric form factor
constitutes an instrument to determine the charge radius,
complementary to atomic spectroscopy methods
\cite{Gao:2021sml}. Possible EMFF sign changes, or
\textit{zero-crossings}, at high $q^2$, are governed by the momentum
dependence of the running strong coupling and mass functions of the
quarks inside the hadron \cite{Segovia:2014aza, Roberts:2020hiw}. Such
zero-crossings are predicted for nucleons by Dyson-Schwinger
calculations but are not yet experimentally established
\cite{Puckett:2017flj}. While neutron data elucidate the $u-d$ quark
flavor dependence \cite{PhysRevLett.106.252003}, the impact of the
$\approx$20 times heavier strange $s$-quark remains rather uncharted
territory. Addressing this question is challenging since the short
lifetime (10$^{-10}$ s) of strange \textit{hyperons} make them virtually
unfeasible for elastic electron-hadron scattering
experiments - the only existing measurement of this kind is for the $\Sigma^-$ hyperon \cite{SELEX:2001fbx}. However, \textit{time-like} EMFFs of hyperons can be accessed
through the process $e^+e^- \to \gamma^* \to Y\bar{Y}$, where $Y$
($\bar{Y}$) denotes a hyperon (antihyperon). Whereas space-like ($q^2$
= $-Q^2 < 0$) EMFFs are real, the time-like ($q^2 > 0$) ones are
complex with a phase reflecting the fact that the virtual photon can
undergo quantum fluctuations into an intermediate multihadron state
\textit{e.g.} $2\pi$ or $3\pi$ \cite{Mangoni:2021qmd}.

For spin 1/2 baryons, the relative phase between the electric and
magnetic form factors manifests itself in transversely polarized final-state
baryons and antibaryons, even for an unpolarized initial state
\cite{Dubnickova:1992ii}. Here, ground-state hyperons have an
advantage over nucleons: their weak, self-analyzing decays offer
straightforward access to their polarization. This was exploited by
BESIII in a pioneering measurement of the EMFF phase of the $\Lambda$
hyperon, obtaining the complete EMFFs at a single energy for the
first time \cite{BESIII:2019nep}. This led to increased activity in
the theory community with a variety of approaches:
$\Lambda\bar{\Lambda}$ final state interactions
\cite{Haidenbauer:2020wyp}, vector meson dominance
\cite{Yang:2019mzq}, covariant spectator model \cite{Ramalho:2019koj}
and dispersive calculations \cite{Lin:2022baj, Mangoni:2021qmd}. In
Ref. \cite{Mangoni:2021qmd}, Mangoni \textit{et al.} outlined a method
to study zero-crossings and even determine the charge root-mean-squared
(RMS) radius $\bar r_E^\Lambda$ of the $\Lambda$ by combining the
BESIII measurement \cite{BESIII:2019nep} with a partial EMFF
determination at two energy points by BaBar
\cite{BaBar:2007fsu}. However, the scarcity of data resulted in a
highly ambiguous solution. Energy-dependent phase measurements are
expected to constrain the calculations and improve the $\Lambda$
charge RMS radius determination.

A dedicated energy scan at center-of-mass
system (CMS) energies between 2.0 and 3.08 GeV was performed in
2014-2015 at BESIII. At these energies, we expect the influence from vector charmonium resonances to be negligible and the production mechanism to be dominated by one-photon exchange. Each energy corresponds to a $q^2$ through the relation $E_{\text{CMS}/}^2/c^2 = q^2$. From the resulting rich data, partial EMFF
information has been obtained for protons \cite{BESIII:2019hdp},
neutrons \cite{BESIII:2021tbq, BESIII:2022rrg}, $\Sigma$
\cite{BESIII:2020uqk, BESIII:2021rkn, BESIII:2023pfv} and $\Xi$
\cite{BESIII:2019cuv, BESIII:2021aer, BESIII:2020ktn}. Furthermore,
exclusive double-tag analyses of the $e^+e^- \to \Lambda\bar{\Lambda}
\to p\pi^-\bar{p}\pi^+$ and $e^+e^- \to \Sigma^+\bar{\Sigma}^- \to
p\pi^0\bar{p}\pi^0$ reactions, enabled the complete
determination of the EMFF for $\Lambda$ at one energy
\cite{BESIII:2019nep} and for $\Sigma^+$ at two energies
\cite{BESIII:2023ynq}. In this Letter, we combine the analysis of
double-tag $e^+e^- \to \Lambda\bar{\Lambda} \to p\pi^-\bar{p}\pi^+$
with single-tag $e^+e^- \to \Lambda X \to p\pi^- X$ and $e^+e^- \to
\bar{\Lambda} X \to \bar{p}\pi^+ X$ at five different energies. This
enables the complete determination of the $\Lambda$ EMFF at five energies within the range 2.3864-3.0800 GeV. In addition, we measure the
production cross-section at nine energies. The integrated luminosities
at all CMS energies are listed in Table \ref{tab:AllCsNums_1}.

The Beijing Spectrometer (BESIII,~\cite{BESIII:2009fln}) at the
Beijing Electron Positron Collider is uniquely suitable for hyperon
EMFF investigations. The BESIII detector covers 93\% of the 4$\pi$
solid angle and is comprised of a small-cell, helium-based multilayer
drift chamber (MDC) for tracking, a time-of-flight (ToF) system based on
plastic scintillators, an electromagnetic calorimeter made of CsI(Tl)
crystals, a muon counter made of resistive plate chambers, and a
superconducting solenoid magnet with a central field of \mbox{1.0 T}.

To determine the modulus of the form factor ratio $R(q^2) =
|G_E(q^2)/G_M(q^2)|$ and the relative phase $\Delta \Phi(q^2) =
\Phi_E(q^2) - \Phi_M(q^2) = {\rm arg}(G_E(q^2)/G_M(q^2))$, we apply
the formalism outlined in Ref. \cite{Faldt:2017kgy}. The double-tag
analysis gives access to five independent angles: the $\Lambda$
scattering angle ($\theta$) in the $e^+e^-$ CMS system, and the
helicity angles $\theta_1$ and $\phi_1$ ($\theta_2$ and $\phi_2$),
defined as the polar and azimuthal angle of the proton (antiproton) in
the helicity frame of the mother $\Lambda$ ($\bar{\Lambda}$). In the
single-tag case, the helicity angles of the untagged hyperon are
integrated out, and by defining the angle $\theta_p$ between the
proton (antiproton) and the normal of the $\Lambda$ ($\bar{\Lambda}$)
scattering plane \cite{Thoren:2022exe}, 
Eqs. 6.55 and 6.56 in Ref. \cite{Faldt:2017kgy} can be simplified to
\begin{linenomath} \begin{align} \begin{split}
    \mathcal{W}(R, &\Delta \Phi, \theta,\theta_p) = 1 + \frac{\tau - R^2}{\tau+R^2} \cos^2\theta \\
    &+ \alpha \sqrt{1 - {\left[\frac{\tau - R^2}{\tau+R^2}\right]}^2}\sin\Delta\Phi\sin\theta\cos\theta\cos\theta_p,
    \label{eq:DecayDistST} \end{split} \end{align} \end{linenomath}

\noindent where $\mathcal{W}(R, \Delta \Phi, \theta,\theta_p)$ is the
joint decay distribution, $\tau = \frac{q^2}{4m^2_\Lambda}$ and
$\alpha$ is the $\Lambda$ decay asymmetry parameter. The latter has
been measured by BESIII \cite{BESIII:2018cnd, BESIII:2021ypr,
  BESIII:2022qax, besiii:2023drj} and CLAS \cite{Ireland:2019uja}. Due
to the numerous new measurements, the Particle Data Group (PDG)
\cite{Workman:2022ynf} continuously updates the world average. For
this work, we adopt the current average $\left<\alpha \right> = (\alpha
-\bar{\alpha})/2$ from the BESIII measurements \cite{BESIII:2018cnd,
  BESIII:2021ypr, BESIII:2022qax, besiii:2023drj}, $\alpha = 0.754$,
assuming CP symmetry \textit{i.e.}  $\alpha = -\bar{\alpha}$. This
allows including measurements of $\bar{\alpha}$ for increased
precision. The resulting value differs from the latest PDG value only
on the sub-percent level, which is well beyond the precision of our
phase measurement.

Detection efficiencies have been obtained from Monte Carlo (MC)
simulations. The ConExc generator, incorporating Initial State
Radiation (ISR) and vacuum polarization (VP) effects \cite{Ping_2014},
has been employed to generate 100 000 signal events at each energy,
including the decay modes $\Lambda \to p\pi^- + c.c.$ and $\Lambda \to
n\pi^0 + c.c.$. The background is studied using the inclusive process
$e^+e^- \to q\bar{q}$ at 2.396 GeV, where the generated 2.86~M events correspond
to an integrated luminosity of 45 nb$^{-1}$. In addition, samples of 100 000
events with the most prominent background channel $e^+e^- \to
\Lambda\bar{\Sigma}^0 + c.c.$ are generated at all CMS energies. The
normalization factors used in the $R(q^2)$ and $\Delta \Phi(q^2)$
parameter estimation are obtained using 5~M phase-space signal events
at each energy. Systematic checks are performed using a BESIII
adaptation of EvtGen~\cite{Lange:2001uf}, describing the full angular
distribution from Ref.~\cite{Faldt:2017kgy}.

In the event selection, the point of closest approach of charged tracks
to the interaction point must be inside a cylindrical volume defined
by the radial distance $V_{xy} \leq 10$ cm and the longitudinal distance $V_{z} \leq 30$ cm from the interaction point. The $z$-axis is
the symmetry axis of the MDC, and the polar angle with respect to the
$z$-axis must fulfill $|\cos\theta_{\text{lab}}| \leq 0.93$.  Furthermore, we
require $p\pi^-$ and $\bar{p}\pi^+$ combinations to have good vertex fits for production and decay, which determine the decay
length $L$ as the distance between the two vertices. The production vertex position, measured independently using Bhabha scattering to a precision of a few hundred $\mu\rm{m}$, is used as a constraint in the vertex fit.
If more than one combination exists in an event, we select the one
giving the largest $L$. This requirement selects the
long-lived $\Lambda/\bar{\Lambda}$ while rejecting random or
non-resonant $p\pi^-$ and $\bar{p}\pi^+$ combinations. The events are
organized into three independent categories, referred to as double-tag
$\Lambda\bar{\Lambda}$, single-tag $\Lambda$ and single-tag
$\bar{\Lambda}$. The double-tag $\Lambda\bar{\Lambda}$ selection is
essentially the same as in Ref.~\cite{BESIII:2019nep}, but with
complete particle identification (PID) of protons, antiprotons and
pions, using ToF and $dE/dx$ information from the MDC. The
$\chi^2_{4C}$ of a four-constraint (4C) kinematic fit, exploiting four-momentum
conservation of the $e^+e^- \to p \pi^- \bar{p} \pi^+$ process, must
be less than 130. Finally, the $p\pi^-$ and $\bar{p}\pi^+$
combinations must satisfy $|M(p\pi) - m_\Lambda| \leq 6$ MeV/$c^2$
which corresponds to $\pm 4 \sigma$ of the peak in the $p\pi$
invariant mass spectrum. These criteria give efficiencies within the
range of $(13-22)\%$ (see $\epsilon_{\Lambda \bar{\Lambda}}$ in Table
\ref{tab:AllCsNums_1}) and a background fraction of less than $1\%$.

In the single-tag selection, we apply the same criteria on $V_{xy}$,
$V_z$, $\cos\theta_{\text{lab}}$, PID and good vertex fits as for double-tag. In addition, we require the decay length significance $L/\Delta L$ to be larger than 0. To exclude events containing $\Lambda$ or $\Lambdabar$ that do not come from the
two-body $e^+e^- \to \Lambda\Lambdabar$ process, we require the
magnitude of the $\Lambda/\Lambdabar$ momentum in the CMS system,
$p_{\Lambda}$, to be within a $\pm$3$\sigma$ window around the predicted
value for each energy; at 2.396 GeV, we require $|p(p\pi) -
p_{\Lambda}| \leq 14.1$ MeV/$c$. Since the background yield is larger
in the single-tag case, we use a tighter invariant mass criterion:
$|M(p\pi) - m_\Lambda| \leq 4.7$ MeV/$c^2$. In the single-tag
$\Lambda$ case, additional background arises due to protons from
secondary interactions with the beam pipe. To eliminate these events,
we require the sum of the $\chi^2$ of the production and decay vertex
fits to satisfy $\chi^2_{\text{prod}} + \chi^2_{\text{decay}} \leq 8$ and $|L/\Delta
L| \leq 30$. Events containing both good $\Lambda \to
p\pi^-$ and $\Lambdabar \to \bar{p}\pi^+$ tags are not included, as they are in the double-tag sample.

For single-tag events, we do not exploit information from the untagged $\Lambda/\Lambdabar$, decaying either \textit{via} the
charged mode or through the neutral $\Lambda \to
n\pi^0/\Lambdabar \to \bar{n}\pi^0$.  The efficiencies for the single-tag
modes, denoted $\epsilon_{1, \Lambda}$, $\epsilon_{2, \Lambda}$,
$\epsilon_{1, \Lambdabar}$ and $\epsilon_{2, \Lambdabar}$, where 1
refers to the charged mode and 2 to the neutral one, are listed in Table \ref{tab:AllCsNums_1}. We note that the efficiencies
$\epsilon_{2, \Lambda}$ and $\epsilon_{2, \Lambdabar}$ are larger
since there is no overlap with the double-tag sample. The background,
estimated using the $M(p\pi^-)$ and $M(\bar{p}\pi^+)$ sidebands (see
\cite{supp}), is found to be $\leq$1.5\% and $\leq$0.5\% for
single-tag $\Lambda$ and single-tag $\Lambdabar$, respectively. The
small background in both double-tag and single-tag data samples is
confirmed in simulations of the inclusive $e^+e^- \to q\bar{q}$
process and reactions with similar topology, \textit{e.g.}  $e^+e^-
\to \Lambda \bar{\Sigma}^0 + c.c.$.

\begin{table*}[t]
\centering
\caption{CMS energies (in GeV), integrated luminosities $\mathcal{L}$
  (in pb$^{-1}$), ISR+VP corrections $(1+\delta)$, background
  subtracted yields $N$, efficiencies $\epsilon$ (in \%) and the
  resulting Born cross-sections (in pb) with statistical and
  systematic uncertainties. For the single-tag efficiencies
  $\epsilon_{1, \Lambda}$, $\epsilon_{2, \Lambda}$, $\epsilon_{1,
    \Lambdabar}$ and $\epsilon_{2, \Lambdabar}$, 1 and 2 refer to the
  charged and neutral decay modes, respectively, of the untagged decay.}
	\label{tab:AllCsNums_1}
    \makebox[\linewidth][c]{%
\begin{ruledtabular}
	\begin{tabular}{c|c|c|c|c|c|c|c|c|c|c|c} 
		Energy & $\mathcal{L}$ & $(1+\delta)$ & $N_{\Lambda \bar{\Lambda}}$ & $N_{\bar{\Lambda}}$ & $N_\Lambda$ & $\epsilon_{\Lambda \Lambdabar}$ & $\epsilon_{1, \Lambdabar}$ & $\epsilon_{2, \Lambdabar}$ & $\epsilon_{1, \Lambda}$ & $\epsilon_{2, \Lambda}$ & $\sigma_{B}$ \\ \hline
		2.3864 & 22.6 &0.98 & $227\pm15$ & $373\pm19$ & $315\pm20$ & $18.6$ & $12.3$ & $31.8$ & $11.7$ & 28.2 &  $132.7(45)(28)$ \\
		2.3960 & 66.9  &0.99 & $596\pm24$ & $973\pm32$ & $575\pm25$ & 18.5 & 12.1 & 31.4 & 7.5 & 18.4 & $120.2(26)(25)$\\
		2.6544 & 67.8  & 1.23 & $224\pm15$ & $300\pm18$ & $244\pm17$ & 21.7 & 9.7 & 31.2 & 9.0 & 27.7 & $31.0(11)(07)$ \\
		2.9000 & 105.5 & 1.57& $117\pm11$ & $120\pm11$ & $108\pm11$ & 19.1 & 6.5 & 23.5 & 6.2 & 22.0 & $8.7(05)(03)$ \\
		2.9500 & 16.0  & 1.65& $16\pm4$ & $12\pm4$ & $14\pm4$ & 17.4 & 6.4 & 23.8 & 6.1 & 21.9 & $7.2(11)(02)$ \\
		2.9810 & 16.0 & 1.71& $16\pm4$ & $13\pm4$ & $13\pm4$ & 18.8 & 5.9 & 22.9 & 5.6 & 21.5 & $7.7(12)(02)$ \\
		3.0000 & 15.8   & 1.75& $13\pm3$ & $12\pm4$ & $14\pm4$ & 12.9 & 6.2 & 23.2 & 5.9 & 21.6 & $7.1(12)(02)$ \\
		3.0200 & 17.3   & 1.75& $11\pm3$ & $12\pm4$ & $14\pm4$ & 14.1 & 6.1 & 23.8 & 5.8 & 22.2 & $5.6(10)(02)$ \\
		3.0800 & 126.6 & 1.85& $65\pm8$ & $60\pm8$ & $62\pm8$ & 14.8 & 4.1 & 16.1 & 3.9 & 15.1 & $4.6(04)(02)$
	\end{tabular}
 \end{ruledtabular}
	}
\end{table*}

The Born cross-section $\sigma_{B}(e^+e^- \to \gamma^* \to
\Lambda\Lambdabar)$ is first calculated for the double-tag
$\Lambda\Lambdabar$, single-tag $\Lambda$ and single-tag $\Lambdabar$ samples separately. For the double-tag case, we use \begin{linenomath}
\begin{align} \label{eq:crossdouble}
	\sigma_{B} = \frac{\sigma_{exp}}{1+\delta} = \frac{N_{\Lambda\Lambdabar}}{\mathcal{L}(1+\delta)\epsilon_{\Lambda\Lambdabar}\mathcal{B}_1^2},
\end{align}    
\end{linenomath}
where $N_{\Lambda\Lambdabar}$ is obtained with the two-dimensional sideband method presented in Ref.
\cite{BESIII:2021ccp}, $\mathcal{L}$ is the integrated luminosity as
given in Table \ref{tab:AllCsNums_1}, $\epsilon_{\Lambda\Lambdabar}$
is the efficiency, $(1+\delta)$ is the ISR and VP correction factor,
and $\mathcal{B}_1=\mathcal{B}(\Lambda \to p
\pi^-)=\mathcal{B}(\Lambdabar \to \bar{p} \pi^+)=63.9\pm0.5\%$ and is
the branching fraction from the PDG. From the single-tag $\Lambda$ sample, $\sigma_{B}$ is calculated using \begin{linenomath}
\begin{align} \label{eq:crosssingle}
	\sigma_{B} = \frac{\sigma_{exp}}{1+\delta} = \frac{N_{\Lambda}}{\mathcal{L} (1+\delta)\mathcal{B}_1\left( \epsilon_{1,\Lambda} \mathcal{B}_1 + \epsilon_{2,\Lambda}\mathcal{B}_2\right)},
\end{align}
\end{linenomath}
where $\mathcal{B}_2=\mathcal{B}(\Lambda \to n \pi^0) =
35.8\pm0.5\%$~\cite{Workman:2022ynf}. The number of single-tag
$\Lambda$ events $N_{\Lambda}$ is determined by sideband subtraction
\cite{supp}. For single-tag $\Lambdabar$, we
replace $N_{\Lambda} \to N_{\Lambdabar}$, $\epsilon_{1,\Lambda} \to
\epsilon_{1,\Lambdabar}$ and $\epsilon_{2,\Lambda} \to
\epsilon_{2,\Lambdabar}$ in Eq. \ref{eq:crosssingle} and assume
$\mathcal{B}(\Lambda \to n \pi^0) = \mathcal{B}(\Lambdabar \to \bar{n}
\pi^0)$.

The ISR and VP correction factors are obtained from ConExc in an
iterative procedure where the input lineshape of the initial step is
obtained by applying a pQCD model \cite{Ping_2016} to data from BaBar
\cite{BaBar:2007fsu} and BESIII \cite{PhysRevD.97.032013}. Applying
the resulting correction factors to our data, we perform a new fit
using a dipole model \cite{Thoren:2022exe}. The resulting refined
lineshape is used as input in the next step. This procedure is
repeated until the difference between the last two iterations becomes
negligible compared to other uncertainties. The
cross-sections obtained for the three independent samples (double-tag
$\Lambda\bar{\Lambda}$, single-tag $\Lambda$ and single-tag
$\Lambdabar$) all agree within the statistical uncertainty and are
combined using the weighted mean. The results are listed in Table
\ref{tab:AllCsNums_1}.

General systematic effects arise from the luminosity (1\% at each
energy point \cite{BESIII:2017lkp}) and $\mathcal{B}_{1,2}$
\cite{Workman:2022ynf}. Systematic uncertainties specific to this
analysis are evaluated for the relatively large samples at 2.396 GeV
and 2.9 GeV \cite{Thoren:2022exe}, while obtained by linear
interpolation or extrapolation in the remaining data points. Our
analysis reveals small but non-negligible effects from differences at
the sub-percent level in the result when varying the
$\chi^2_{\rm{4C}}$ (double-tag) and $\Lambda$/$\Lambdabar$ momentum
(single-tag) selection requirements. In addition, there is a small
uncertainty of (0.1-0.8)\% in the $\Lambda$/$\Lambdabar$
reconstruction efficiency and above 2.9 GeV, a contribution of
(0.5-1.2)\% from the $\Lambda\bar{\Sigma}^0$ background.  The uncertainty from the ISR and VP corrections were evaluated by using a pQCD fit for the cross-section lineshape. The difference between the dipole and pQCD model results, (0.1-0.6)\%, is quoted as the uncertainty. Since the MC
data used for the efficiency correction are generated with an angular
distribution governed by $R$, the uncertainty in the input parameter $R$
will propagate to the cross-section.  For energies above 2.9 GeV, this
uncertainty is fairly large \textit{i.e.} (2.2-2.8)\%, while only
within (0.6-1.4)\% below 2.9 GeV. The total systematic uncertainty is
well below the statistical uncertainty in all data points except at
2.396 GeV, where it is of similar magnitude, see Table
\ref{tab:AllCsNums_1}.

\begin{figure}[t]
    \centering
     \includegraphics[scale=0.43, trim={2cm 0 2cm 1cm,clip}]{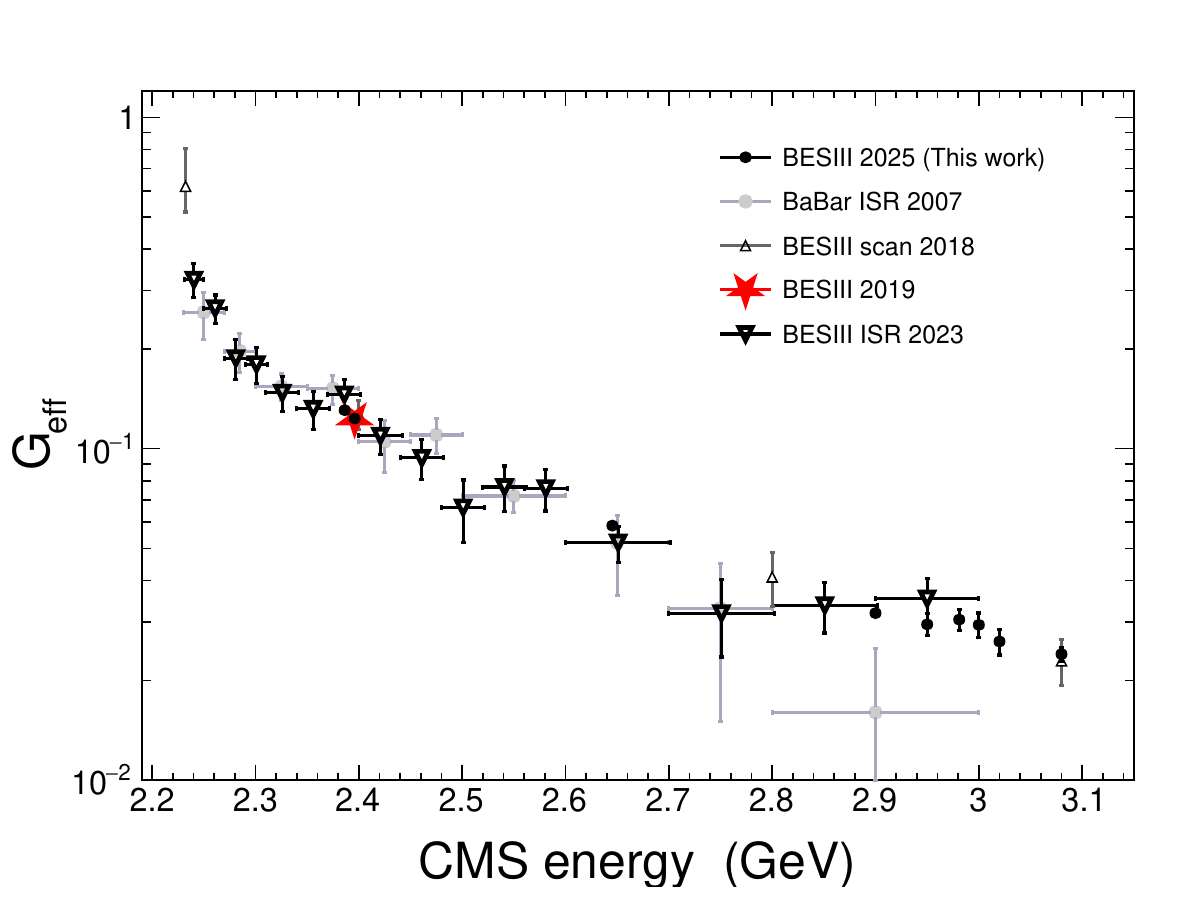}
	\caption{The effective form factor $G_{\text{eff}}$ from this work (black dots), BESIII ISR \cite{BESIII:2023ioy}, BESIII energy scan \cite{BESIII:2017hyw,BESIII:2019nep} and BaBar ISR \cite{BaBar:2007fsu}. The horizontal error bars denote the bin size. 
    }
    \label{fig:gEMFFs}
\end{figure}

Assuming that the dominant process is one-photon exchange
($e^+e^-\rightarrow\gamma^*\rightarrow \Lambda\bar{\Lambda}$), the
effective form factor $G_{\text{eff}}(q^2)$ can be calculated:
\begin{linenomath}
 \begin{equation}
\begin{split}
|G_{\text{eff}}(q^2)| \equiv &\sqrt{\frac{\sigma_B(q^2)}{(1+\frac{1}{2\tau})(\frac{4\pi\alpha^2_{\text{EM}}\beta }{3q^2})}}. 
\label{equ-effectiveff}
\end{split}
\end{equation}   
\end{linenomath}
\noindent Here, $\alpha_{\text{EM}}$ is the fine-structure constant,
$\beta=\sqrt{1-4m^2_{\Lambda}/q^2}$ the $\Lambda$ velocity,
$m_{\Lambda}$ the $\Lambda$ mass, and $\tau =
q^2/(4m^2_{\Lambda})$. The results are presented in Table
\ref{tab:AllResSum}, shown in Fig. \ref{fig:gEMFFs} and have
significantly improved precision compared to previous measurements
\cite{BESIII:2023ioy}.

\begin{figure}[t]
    \centering
    \hspace{-2mm}
    \includegraphics[trim={10mm 10mm 0mm 10mm,clip},scale=0.45]{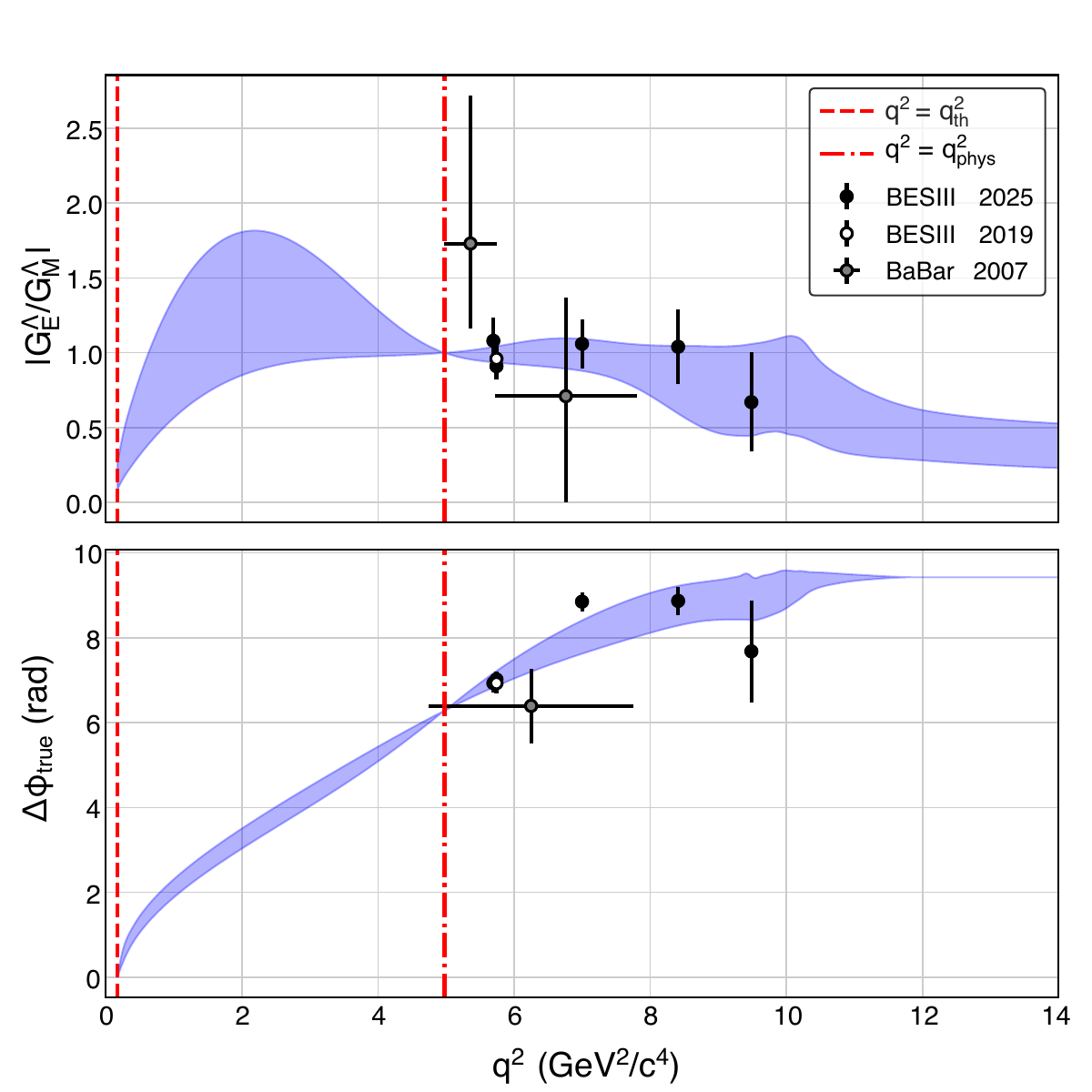}
	\caption{Top: The modulus $R=|G_E/G_M|$ from this work (black dots), from a previous BESIII study (white dot) \cite{BESIII:2019nep} and from BaBar (gray dots) \cite{BaBar:2007fsu}. The band represents a theoretical analysis outlined in Ref. \cite{Mangoni:2021qmd}, showing the most probable scenario $[N_{\text{th}},N_{\text{asy}}] = [0,3]$. The BESIII 2019 data point is superseded by the point at the same CMS energy in this work and is therefore not included in the fit. The red dashed line marks the theoretical threshold $q^2_{th}$ and the red dashed-dotted line the physical threshold $q^2_{\text{phys}}$. Bottom: The corresponding measurements of the relative phase $\Delta\Phi$. }
    \label{fig:EMFFs}
\end{figure}

The form factor ratio $R=R(q^2)$ and phase $\Delta \Phi=\Delta
\Phi(q^2)$ are obtained at the five energies with sufficient data. We use a maximum log-likelihood (MLL) fit, where
the likelihood function combines the double-tag, single-tag and
background cases (for details, see
Refs.\cite{supp,Thoren:2022exe}). In addition, separate fits are carried out
for each sub-sample. We perform bias tests using an
ensemble of 1000 MC samples, generated by taking the BESIII detector
response and background into account. Each MC sample has the
same size as the real data samples. The MLL estimator is found to be
unbiased for both $R$ and $\Delta\Phi$ at all
sub-samples at all energies, except for $\Delta\Phi$ at 2.396 GeV. A
detailed MC study found that for 10\% of the single-tag
$\Lambda$ samples, the likelihood function does not have a clear
maximum and instead, the MLL fitter assigns a phase $\Delta \Phi$ =
90$^\mathrm{o}$ regardless of the input value. The likelihood function
obtained from the real single-tag $\Lambda$ sample at 2.396 GeV was
found to have the same peculiar shape, in contrast to that of all
other samples. Therefore, the $\Delta \Phi$ at this energy is
determined only from double-tag $\Lambda\Lambdabar$ and single-tag
$\Lambdabar$ events. The $R$ and $\Delta\Phi$ at all other
energies, are determined using all three samples.  The results are shown in Table
\ref{tab:AllResSum} and Fig. \ref{fig:EMFFs}.

A surprising result is that while $R$ is fairly constant,
the phase $\Delta \Phi$ increases by about 90$^\mathrm{o}$ between
2.396 GeV and 2.645 GeV. We perform an independent test where we
exploit the $\Lambda\Lambdabar$ spin correlations $C_{\bar{x}z} =
C_{\bar{z}x}$ \cite{PANDA:2020zwv}, which depend explicitly on
$\cos\Delta\Phi$. By calculating these correlations in each hemisphere
of the $\Lambda$ scattering angle and comparing to the prediction for
each phase hypothesis, we confirm that $\Delta \Phi$ indeed is below
$90^{\mathrm{o}}$ at 2.396 GeV and above $90^{\mathrm{o}}$ at 2.645
GeV and 2.9 GeV.

Systematic effects are investigated by varying all selection criteria
for all samples. In most cases the effects are negligible, except for
the $M(p\pi)$ window that slightly influences $\Delta \Phi$, and the requirement on the $\chi^2$ from the vertex fit
that has a small impact on $R$. In addition, there
is an uncertainty in the $\Lambda$/$\Lambdabar$ reconstruction
efficiency for both $R$ and $\Delta \Phi$. Background effects
are studied by sampling MC data mimicking the sideband events in
real data. These background samples are added to signal MC events in
1000 MLL fits. We find the effect to be small, though
slightly larger for $R$ than for $\Delta \Phi$. The systematic
uncertainties of $R$ and $\Delta\Phi$ are summarized in Table
\ref{tab:SystSumDPhi}.

\begin{table}[t]
    \centering
    \caption{The measured $G_{\text{eff}}$, the ratio $R=|G_E/G_M|$, and the relative phase $\Delta\Phi$ at each CMS energy (GeV). The first uncertainty is statistical and the second systematic.}
    \begin{tabular}{c|c|c|c}
	    \hline \hline
    Energy &  $G_{eff}$  & $R$ & $\Delta\Phi$ ($^\circ$)\\ 
    \hline
     
    2.3864     & $0.1307(22)(14)$ & $1.08(15)(03)$ & $37(12)(01)$ \\
    2.3960       & $0.1233(13)(13)$ & $0.91(08)(02)$ & $42(10)(02)$\\
    2.6454        & $0.0586(10)(07)$ & $1.06(16)(03)$&  $147(13)(01)$\\
    2.9000         & $0.0319(09)(06)$ & $1.04(25)(03)$&  $148(19)(01)$\\
	    2.9500 & $0.0295(22)(04)$ & - & - \\
    2.9810 & $0.0305(22)(04)$ & - & - \\
    3.0000 & $0.0294(25)(04)$ & - & - \\
    3.0200 & $0.0262(23)(05)$ & - & - \\
    3.0800 & $0.0240(10)(05)$ & $0.67(33)(03)$&  $80(69)(02)$\\ \hline \hline
    \end{tabular}
    \label{tab:AllResSum}
\end{table}

\begin{table}[b]
    \centering
    \caption{Systematic uncertainties in $R$ ($\Delta\Phi$) in \% at
      each CMS energy (GeV). The sources are from (I) the $M(p\pi)$
      window, (II) vertex fit $\chi^2$ requirement, (III)
      $\Lambda/\bar{\Lambda}$ reconstruction efficiency, and (IV)
      background effects.}
    \label{tab:SystSumDPhi}
    \makebox[\linewidth][c]{%
    \begin{tabular}{l|l|l|l|l|l}
            \hline \hline
    Source     & 2.3864 & 2.3960 & 2.6454  & 2.9000  & 3.0800  \\ \hline
            (I) & 0 (3.7) & 0 (3.6) & 0 (1.9) & 0 (0) & 0 (0) \\
            (II) & 1.6 (0) & 1.6 (0) & 0.8 (0) & 0 (0) & 0 (0)\\
            (III) & 1.8 (0.5) & 0.7 (0.6) &  1.9 (0.2)  & 1.9 (0.1) & 4.5 (2.6) \\ 
            (IV) & 1.6 (0.6) & 1.6 (0.6) & 1.8 (1.0)  & 2.0 (1.5) & 2.1 (1.8) \\ \hline
            Total & 2.9 (3.8) & 2.4 (3.7) & 2.7 (2.2) & 2.8 (1.5) & 5.0 (3.2) \\ \hline \hline
    \end{tabular}}
\end{table}

The true phase can be the quoted value plus any integer multiple of
$2\pi$, \textit{i.e.} $\Delta \Phi_{\text{true}} = \Delta \Phi +
N\cdot2\pi$. However, the dispersive approach in
Ref. \cite{Mangoni:2021qmd} can help to resolve this
ambiguity. Calculations are performed for different scenarios for the
quantities \begin{align}
    N_{\text{th}} = \frac{1}{\pi}arg(\frac{G_E(q^2_{\text{th}})}{G_M(q^2_{\text{th}})}),
    N_{\text{asy}} = \frac{1}{\pi}arg(\frac{G_E(q^2_{\text{asy}})}{G_M(q^2_{\text{asy}})}),
    \label{nth}
\end{align}
\noindent where $q^2_{\text{th}}$ corresponds to the squared-mass of the
lowest-lying hadronic state that can couple to the virtual photon and
the $\Lambda\Lambdabar$ final state, \textit{i.e.} $\pi^+\pi^-\pi^0$,
and $q^2_{\text{asy}}$ is the scale where the phase $\Delta\Phi(q^2)$
approaches an integer multiple of $\pi$. At the physical threshold
$q^2_{\text{phys}} = 4m_{\Lambda}^2$, only $s$-waves can contribute to the
$\Lambda\Lambdabar$ production and therefore $R(q^2_{\text{phys}}) \mbeq 1$
and $\Delta\Phi(q^2_{\text{phys}}) \mbeq 2k\pi$. The number of zero-crossings of
the complex ratio $G_E(q^2)/G_M(q^2)$ is given by $\Delta N =
N_{\text{asy}}-N_{\text{th}}$, following Cauchy's argument principle. In the calculations, the imaginary part of the ratio
is parameterized by Chebyshev polynomials of the first kind. Seven or
eight degrees appear optimal and give consistent results as shown in
Table \ref{tab:chargerad}. The $\chi^2/\nu$ is obtained from the best fit result. To calculate the probability, we perform simulations of pseudo-experiments of different $[N_{\text{th}},N_{\text{asy}}]$ scenarios, using the measured values and their uncertainties as input. The probability $P$ is given by the relative occurrence of a $[N_{\text{th}},N_{\text{asy}}]$ scenario for a large number of pseudo-experiments. The most probable scenario,
$[N_{\text{th}},N_{\text{asy}}] = [0,3]$ ($P=85\%)$ shown in Fig.~\ref{fig:EMFFs},
is strikingly more probable than the second scenario $[N_{\text{th}},N_{\text{asy}}]
= [0,4]$. Notably, all possible solutions imply multiple
zero-crossings of $G_E(q^2)/G_M(q^2)$, in line with
Dyson-Schwinger-based predictions for nucleons \cite{Segovia:2014aza}
and vector mesons \cite{Xu:2019ilh}.

The sign-normalized RMS charge radius is defined as 
\begin{equation}
    \bar r_E^\Lambda=\text{sgn}\left( \langle r^\Lambda_E \rangle^2 \right) \sqrt{\left| \langle r^\Lambda_E \rangle^2 \right|}.
\end{equation}
\noindent For neutral baryons with an anomalous magnetic moment, $\langle r^\Lambda_E \rangle^2$ can be calculated from the derivative of $R(q^2)$ at $q^2 = 0$  \cite{Mangoni:2021qmd}. The results are presented in Table \ref{tab:chargerad}, including the probability of each solution as calculated using Monte Carlo simulations \cite{Mangoni:2021qmd}. The most probable $[N_{\text{th}},N_{\text{asy}}]$ scenario indicates a negative RMS charge radius ($\bar{r}_E^{\Lambda} = -0.076 \pm 0.043$ fm for a seven degree polynomial). The effect is $\approx$2 standard deviations and the conclusion is valid also when lower degree polynomials are used in the fit. This indicates that the negative $sd$ quark pair in the $\Lambda$ is close to the center of the $\Lambda$, similar to the $dd$ pair in the neutron \cite{Atac:2021wqj}. In summary, our measurements of time-like EMFFs, combined with dispersive relations, open new avenues to study the properties of unstable matter containing strange quarks and beyond.

\begin{table}[t]
    \centering
    \caption{The RMS charge radius $\bar{r}_E^{\Lambda}$ (in fm) and the
      probability $P$ (in \%) for different $(N_{\text{th}}, N_{\text{asy}})$
      scenarios, obtained by fitting Chebyshev polynomials of the 6th, 7th and 8th order ~\cite{Mangoni:2021qmd} to data from this work and
      Ref.~\cite{BaBar:2007fsu}.}
    \label{tab:chargerad}
    \begin{tabular}{c|c|c|c|c|c|c}
    \hline
    \hline
      & \multicolumn{2}{|c|}{6th, $\chi^2/\nu = 2.5$} & \multicolumn{2}{|c|}{7th, $\chi^2/\nu = 1.9$} & \multicolumn{2}{|c|}{8th, $\chi^2/\nu = 1.9$}\\
     \hline
     $(N_{\text{th}},N_{\text{asy}})$  & $\bar{r}_E^{\Lambda}$ & $P$ & $\bar{r}_E^{\Lambda}$ & $P$ & $\bar{r}_E^{\Lambda}$ & $P$ \\ \hline
    (0,3) & -0.125 (70) & 57 & -0.076 (43) & 83 & -0.077 (40) & 85 \\
    (0,4) & -0.109 (62) & 26 & -0.040 (30) & 12 & -0.040 (30) & 7.6 \\
    (1,3) & 0.070 (30) & 8.0 & 0.031 (06) & 3.0 & 0.055 (28) & 3.8 \\
    (-1,1) & 0.152 (59) & 5.0 & 0.025 (19) & 1.0 & 0.019 (16) & 1.0 \\
             \hline 
             \hline 
    \end{tabular}
\end{table}

\begin{acknowledgments}

The BESIII Collaboration thanks the staff of BEPCII (https://cstr.cn/31109.02.BEPC) and the IHEP computing center for their strong support. This work is supported in part by National Key R\&D Program of China under Contracts Nos. 2020YFA0406300, 2020YFA0406400, 2023YFA1606000; National Natural Science Foundation of China (NSFC) under Contracts Nos. 11635010, 11735014, 11935015, 11935016, 11935018, 12025502, 12035009, 12035013, 12061131003, 12192260, 12192261, 12192262, 12192263, 12192264, 12192265, 12221005, 12225509, 12235017, 12361141819; the Chinese Academy of Sciences (CAS) Large-Scale Scientific Facility Program; the CAS Center for Excellence in Particle Physics (CCEPP); Joint Large-Scale Scientific Facility Funds of the NSFC and CAS under Contract No. U1832207; CAS under Contract No. YSBR-101; 100 Talents Program of CAS; The Institute of Nuclear and Particle Physics (INPAC) and Shanghai Key Laboratory for Particle Physics and Cosmology; Agencia Nacional de Investigación y Desarrollo de Chile (ANID), Chile under Contract No. ANID PIA/APOYO AFB230003; German Research Foundation DFG under Contract No. FOR5327; Istituto Nazionale di Fisica Nucleare, Italy; Knut and Alice Wallenberg Foundation under Contracts Nos. 2021.0174, 2021.0299; Ministry of Development of Turkey under Contract No. DPT2006K-120470; National Research Foundation of Korea under Contract No. NRF-2022R1A2C1092335; National Science and Technology fund of Mongolia; National Science Research and Innovation Fund (NSRF) via the Program Management Unit for Human Resources \& Institutional Development, Research and Innovation of Thailand under Contract No. B50G670107; Polish National Science Centre under Contract No. 2019/35/O/ST2/02907; Swedish Research Council under Contract Nos. 2019.04595 and 2021.04567; The Swedish Foundation for International Cooperation in Research and Higher Education under Contract No. CH2018-7756; U. S. Department of Energy under Contract No. DE-FG02-05ER41374


\end{acknowledgments}

\appendix

\
\nocite{*}

\bibliography{apssamp}

\end{document}